\documentstyle[12pt]{article}

\catcode`\@=11
\def\@citex[#1]#2{%
\if@filesw \immediate \write \@auxout {\string \citation {#2}}\fi
\@tempcntb\m@ne \let\@h@ld\relax \def\@citea{}%
\@cite{%
  \@for \@citeb:=#2\do {%
    \@ifundefined {b@\@citeb}%
      {\@h@ld\@citea\@tempcntb\m@ne{\bf ?}%
      \@warning {Citation `\@citeb ' on page \thepage \space undefined}}%
      {\@tempcnta\@tempcntb \advance\@tempcnta\@ne%
      \@tempcntb\number\csname b@\@citeb \endcsname \relax%
      \ifnum\@tempcnta=\@tempcntb 
	\ifx\@h@ld\relax%
	  \edef \@h@ld{\@citea\csname b@\@citeb\endcsname}%
	\else%
	  \edef\@h@ld{\ifmmode{-}\else--\fi\csname b@\@citeb\endcsname}%
	\fi%
      \else
	\@h@ld\@citea\csname b@\@citeb \endcsname%
	\let\@h@ld\relax%
      \fi}%
    \def\@citea{,\penalty\@highpenalty\,}%
  }\@h@ld
}{#1}}

\def\@citeb#1#2{{[#1]\if@tempswa , #2\fi}}
\def\@citeu#1#2{{$^{#1}$\if@tempswa , #2\fi }}
\def\@citep#1#2{{#1\if@tempswa , #2\fi}}

\def\bcites{         
	\catcode`\@=11
	\let\@cite=\@citeb
	\catcode`\@=12
}

\def\upcites{         
	\catcode`\@=11
	\let\@cite=\@citeu
	\catcode`\@=12
}

\def\plaincites{      
	\catcode`\@=11
	\let\@cite=\@citep
	\catcode`\@=12
}

\newcount\hour
\newcount\minute
\newtoks\amorpm
\hour=\time\divide\hour by 60
\minute=\time{\multiply\hour by 60 \global\advance\minute by-\hour}
\edef\standardtime{{\ifnum\hour<12 \global\amorpm={am}%
	\else\global\amorpm={pm}\advance\hour by-12 \fi
	\ifnum\hour=0 \hour=12 \fi
	\number\hour:\ifnum\minute<10 0\fi\number\minute\the\amorpm}}
\edef\militarytime{\number\hour:\ifnum\minute<10 0\fi\number\minute}

\def\draftlabel#1{{\@bsphack\if@filesw {\let\thepage\relax
   \xdef\@gtempa{\write\@auxout{\string
      \newlabel{#1}{{\@currentlabel}{\thepage}}}}}\@gtempa
   \if@nobreak \ifvmode\nobreak\fi\fi\fi\@esphack}
	\gdef\@eqnlabel{#1}}
\def\@eqnlabel{}
\def\@vacuum{}
\def\marginnote#1{}
\def\draftmarginnote#1{\marginpar{\raggedright\scriptsize\tt#1}}
\def\draft{
	\pagestyle{plain}
	\overfullrule=2pt
	\oddsidemargin -.5truein
	\def\@oddhead{\sl \phantom{\today\quad\militarytime} \hfil
	\smash{\Large\sl DRAFT} \hfil \today\quad\militarytime}
	\let\@evenhead\@oddhead
	\let\label=\draftlabel
	\let\marginnote=\draftmarginnote
	\def\ps@empty{\let\@mkboth\@gobbletwo
	\def\@oddfoot{\hfil \smash{\Large\sl DRAFT} \hfil}
	\let\@evenfoot\@oddhead}
	\def\@eqnnum{(\theequation)\rlap{\kern\marginparsep\tt\@eqnlabel}%
	\global\let\@eqnlabel\@vacuum}  }

\def\blackfonts{
	\font\blackboard=msbm10 scaled\magstep1
	\font\blackboards=msbm8
	\font\blackboardss=msbm6
}

\def\nblack{            
	\def\ZZ{{Z \n{10} Z}}
	\def\NN{{N \n{14} N}}
	\def\CC{{C \n{11} C}}
	\def\RR{{R \n{11} R}}
	\def\QQ{{Q \n{12} Q}}
	\def\PP{{P \n{11} P}}
}

\def\prep{         
	\catcode`\@=11
	\input art10.sty
	\catcode`\@=12
	
	\let\small\null
	\def\blackfonts{
		\font\blackboard=msbm10
		\font\blackboards=msbm7
		\font\blackboardss=msbm5
	}
	\let\sl\it
	\twocolumn
	\sloppy
	\voffset=-2.54truecm
	\hoffset=-2.54truecm
	\flushbottom
	\parindent 1em
	\leftmargini 2em
	\leftmarginv .5em
	\leftmarginvi .5em
	\marginparwidth 48pt
	\marginparsep 10pt
	\setlength{\columnsep}{2truecm}
	\setlength{\textwidth}{25.4truecm}
	\setlength{\textheight}{17truecm}
	\baselineskip=16pt
	\oddsidemargin .18truein
	\evensidemargin .17truein
}

\def\eqalign#1{\null\,\vcenter{\openup\jot\m@th
  \ialign{\strut\hfil$\displaystyle{##}$&$\displaystyle{{}##}$\hfil
      \crcr#1\crcr}}\,}
\def\eqalignno#1{\displ@y \tabskip\centering
  \halign to\displaywidth{\hfil$\@lign\displaystyle{##}$\tabskip\z@skip
    &$\@lign\displaystyle{{}##}$\hfil\tabskip\centering
    &\llap{$\@lign##$}\tabskip\z@skip\crcr
    #1\crcr}}

\def\section{\@startsection {section}{1}{\z@}{3.ex plus 1ex minus
 .2ex}{2.ex plus .2ex}{\large\bf}}
\def\subsection{\@startsection{subsection}{2}{\z@}{2.75ex plus 1ex minus
 .2ex}{1.5ex plus .2ex}{\bf}}

\def\appendix{{\newpage\section*{Appendix}}\let\appendix\section%
	{\setcounter{section}{0}
	\gdef\thesection{\Alph{section}}}\section}

\def\abstract{\if@twocolumn
\section*{Abstract}
\else 
\begin{center}
{\bf Abstract\vspace{-.5em}\vspace{0pt}}
\end{center}
\quotation
\fi}

\catcode`\@=12

\def\noj#1,#2,{{\bf #1} (19#2)\ }
\def\jou#1,#2,#3,{{\sl #1\/ }{\bf #2} (19#3)\ }
\def\ann#1,#2,{{\sl Ann.\ Physics\/ }{\bf #1} (19#2)\ }
\def\cmp#1,#2,{{\sl Comm.\ Math.\ Phys.\/ }{\bf #1} (19#2)\ }
\def\invm#1,#2,{{\sl Invent.\ Math.\/ }{\bf #1} (19#2)\ }
\def\cq#1,#2,{{\sl Class.\ Quantum Grav.\/ }{\bf #1} (19#2)\ }
\def\cqg#1,#2,{{\sl Class.\ Quantum Grav.\/ }{\bf #1} (19#2)\ }
\def\ijmp#1,#2,{{\sl Int.\ J.\ Mod.\ Phys.\/ }{\bf A#1} (19#2)\ }
\def\jmphy#1,#2,{{\sl J.\ Geom.\ Phys.\/ }{\bf #1} (19#2)\ }
\def\jams#1,#2,{{\sl J.\ Amer.\ Math.\ Soc.\/ }{\bf #1} (19#2)\ }
\def\grg#1,#2,{{\sl Gen.\ Rel.\ Grav.\/ }{\bf #1} (19#2)\ }
\def\mpl#1,#2,{{\sl Mod.\ Phys.\ Lett.\/ }{\bf A#1} (19#2)\ }
\def\nc#1,#2,{{\sl Nuovo Cim.\/ }{\bf #1} (19#2)\ }
\def\np#1,#2,{{\sl Nucl.\ Phys.\/ }{\bf B#1} (19#2)\ }
\def\pl#1,#2,{{\sl Phys.\ Lett.\/ }{\bf #1B} (19#2)\ }
\def\pla#1,#2,{{\sl Phys.\ Lett.\/ }{\bf #1A} (19#2)\ }
\def\pr#1,#2,{{\sl Phys.\ Rev.\/ }{\bf #1} (19#2)\ }
\def\prd#1,#2,{{\sl Phys.\ Rev.\/ }{\bf D#1} (19#2)\ }
\def\prl#1,#2,{{\sl Phys.\ Rev.\ Lett.\/ }{\bf #1} (19#2)\ }
\def\prp#1,#2,{{\sl Phys.\ Rept.\/ }{\bf #1C} (19#2)\ }
\def\ptp#1,#2,{{\sl Prog.\ Theor.\ Phys.\/ }{\bf #1} (19#2)\ }
\def\ptpsup#1,#2,{{\sl Prog.\ Theor.\ Phys.\/ Suppl.\/ }{\bf #1} (19#2)\ }
\def\rmp#1,#2,{{\sl Rev.\ Mod.\ Phys.\/ }{\bf #1} (19#2)\ }
\def\yadfiz#1,#2,#3[#4,#5]{{\sl Yad.\ Fiz.\/ }{\bf #1} (19#2) #3%
\ [{\sl Sov.\ J.\ Nucl.\ Phys.\/ }{\bf #4} (19#2) #5]}
\def\zh#1,#2,#3[#4,#5]{{\sl Zh.\ Exp.\ Theor.\ Fiz.\/ }{\bf #1} (19#2) #3%
\ [{\sl Sov.\ Phys.\ JETP\/ }{\bf #4} (19#2) #5]}

\def\beq{\begin{equation}}
\def\eeq{\end{equation}}
\def\beqar{\begin{eqnarray}}
\def\eeqar{\end{eqnarray}}

\def\nfrac#1#2{{\displaystyle{\vphantom1\smash{\lower.5ex\hbox{\small$#1$}}%
	\over\vphantom1\smash{\raise.25ex\hbox{\small$#2$}}}}}

\def\p#1{\mskip#1mu}
\def\n#1{\mskip-#1mu}
\def\stop{\p6.}
\def\comma{\p6,}

\def\lae{\mathrel{\mathop{\smash{\lower .5 ex \hbox{$\stackrel<\sim$}}}}}
\def\lae{\mathrel{\mathop{\smash{\lower .5 ex \hbox{$\stackrel>\sim$}}}}}

\def\pa{\partial}

\def\l:{\mathopen{:}\,}
\def\r:{\,\mathclose{:}}

\def\[{\left[}          \def\]{\right]}
\def\({\left(}          \def\){\right)}
\def\<{\left<}          \def\>{\right>}

\catcode`\@=11
\def\theequation{\arabic{equation}}
\catcode`\@=12

\nblack
\bcites

\nblack

\catcode`\@=11
\def\theequation{\thesection.\arabic{equation}}
\@addtoreset{equation}{section}
\@addtoreset{footnote}{section}
\@addtoreset{footnote}{subsection}
\catcode`\@=12

\newcommand{\beqn}{\begin{equation}}
\newcommand{\eeqn}{\end{equation}}
\newcommand{\beqnarray}{\begin{eqnarray}}
\newcommand{\eeqnarray}{\end{eqnarray}}
\newcommand {\nono} {\nonumber \\} 

\newcommand {\ul} [1] {\underline {#1}}

\newcommand {\omy} {\mbox{$\Omega$}}

\newcommand {\tofl} {\mbox{$e^{i\phi_L}$}}
\newcommand {\tofr} {\mbox{$e^{i\phi_R}$}}

\newcommand {\qeq} [1] {(\ref {eq:#1})}
\begin{document}
\begin{titlepage}

\begin{center}
June 18, 1996
\hfill LBNL-038974, UCB-PTH-96/26 \\
\hfill                  WIS-96/24/June-PH               \\
\hfill                  hep-th/9606112

\vskip 1 cm
{\large \bf  D-Branes on  Calabi-Yau Spaces and Their Mirrors 
\\}
\vskip 1 cm 
{Hirosi Ooguri, Yaron Oz
\footnote{Permanent address:
Department of Particle Physics,
Weizmann Institute of Science,
 76100 Rehovot Israel.}
and Zheng Yin }\\
\vskip 0.5cm
{\sl Department of Physics,
University of California at Berkeley\\
366 Le\thinspace Conte Hall, Berkeley, CA 94720-7300, U.S.A.\\
and\\
Theoretical Physics Group, Mail Stop 50A--5101\\
Lawrence Berkeley National Laboratory, 
Berkeley, CA 94720, U.S.A.\\}

\end{center}

\vskip 0.5 cm
\begin{abstract}
We study the boundary states of D-branes wrapped around 
supersymmetric cycles in a general Calabi-Yau manifold. 
In particular, we show how the geometric data on the cycles 
are encoded in the boundary states. As an application, we 
analyze how the mirror symmetry transforms D-branes, and 
we verify that it is consistent with the conjectured 
periodicity and the monodromy of the Ramond-Ramond field 
configuration on a Calabi-Yau manifold. This also enables 
us to study open string worldsheet instanton corrections 
and relate them to closed string instanton counting. 
The cases when the mirror symmetry is realized as T-duality 
are also discussed.

\end{abstract}

\end{titlepage}

\section{Introduction}
D-branes in type II string theories 
have been identified as Ramond-Ramond charged
BPS states \cite{POL}. In the presence 
of a D-brane, the boundary conditions
for open strings are modified in such a way
that Dirichlet boundary conditions
are allowed in addition to the Neumann 
boundary conditions.
The study of D-branes and its applications has 
been mainly restricted
to the cases where the D-brane worldvolume is flat.
In \cite{bsv2}, a study of D-branes wrapped on 
curved spaces has been 
carried out in the long wavelength limit.

In this paper we will present a framework at the SCFT level
for the study of D-branes on Calabi-Yau spaces.
Perturbative string computations in the presence of
a D-brane can be formulated by using a boundary
state which describes how closed strings are
emitted or absorbed on the D-brane worldvolume. 
In the case of the fully Neumann boundary condition
near the flat background,
the boundary state was constructed in \cite{boundary}.
Our main object of study is the boundary state for a 
D-brane wrapping on a non-trivial supersymmetric cycle 
in a Calabi-Yau space. In particular, we examine how
the geometric data on the cycle are encoded in the
boundary state.

The analysis of the boundary state will enable us to find the way mirror
symmetry transforms D-brane configurations. 
It has been observed that, for a Calabi-Yau 3-fold $M$,
the mirror symmetry not only maps the even cohomology of $M$
to the odd cohomology of its mirror $\widetilde{M}$, but
it does so while respecting the integral structure
of the cohomologies \cite{al}.  
Based on this, it was conjectured by Aspinwall and Morrison
\cite{AM} that the Ramond-Ramond field on a Calabi-Yau
space must have a certain periodicity reflecting this
integral structure. This way, the mirror map can be extended 
to the Ramond-Ramond field configurations.  
We will verify that this conjecture is consistent with 
the mirror map between D-brane configurations.  

The precise understanding of the mirror symmetry between
D-branes enables us to study open string worldsheet instanton effects.
We will find that the chiral primary part of the 
boundary states for 0, 2 and 3-cycles in a Calabi-Yau 3-fold does
not receive instanton corrections
while the instanton corrections for 4 and 6-cycles
can be expressed in term of the closed string worldsheet instantons
on the same manifold. 

The paper is organized as follows: In section 2 we classify 
boundary conditions for $N=2$ SCFT which preserves half of 
the spacetime supersymmetry and the $N=1$ worldsheet supersymmetry. 
We then examine how these boundary conditions are realized  
by D-branes wrapping on cycles in a Calabi-Yau manifold. One may regard this
as a microscopic version of the analysis by K.~Becker, M.~Becker and
Strominger \cite{BBS}, where they studied the condition on the 
supersymmetric cycles
using the low-energy effective actions for $p$-branes. 
In section 3 we will study the algebraic and geometrical 
structures of the boundary states of D-branes wrapped on supersymmetric
cycles in Calabi-Yau spaces. We will distinguish between the middle-dimensional
and even-dimensional cycles,
and find the dependence of the boundary states on the choice of
the cycles as well as the complex and  
K\"ahler moduli of the Calabi-Yau
space.
Section 4 will be devoted to the analysis of the mirror transformation of
D-brane configurations. 
In section 5 we will present examples where the mirror symmetry
is realized as T-duality on tori and Calabi-Yau
orbifolds. 
Section 6 will be devoted to a discussion.
In the appendix we discuss the construction of boundary states for
Gepner models. We present an example that exhibits the relation between
the boundary conditions for the model and the supersymmetric cycles.

\section{Supersymmetric cycles in Calabi-Yau manifolds}
In this section we will classify 
the boundary conditions for $N=2$ SCFT which preserves half of 
the spacetime supersymmetry and the $N=1$ worldsheet supersymmetry. 
We will then examine how these boundary conditions are realized  
by D-branes wrapping on cycles in a Calabi-Yau manifold.
Here we will consider the case when the sigma-model for
the Calabi-Yau manifold has one set of $N=2$ superconformal
algebra for the left-movers and one set for the right movers. 
It is straightforward to extend this analysis to the case where
we have more than one set of $N=2$ algebras, such as 
$T^{2d}$ with $d \geq 2$.

\subsection{Boundary conditions for $N=2$ SCFT}

The supersymmetric sigma-model for a Calabi-Yau manifolds has $N=2$
superconformal algebra (SCA). Throughout this paper, we set the signs 
of the left and the right $U(1)$ currents to be
\beq \label {eq:n2j_d}
	J_L  =  g_{i\bar{j}} \psi_L^i \psi_L^{\bar{j}},~~~~~         
	J_R  =  g_{i\bar{j}} \psi_R^i  \psi_R^{\bar{j}}\comma
\eeq
which determines the convention for $G^{\pm}$  as
\beqar \label {eq:n2g_d}
	G_L^+ & = & g_{i\bar{j}} \psi_L^i  \pa X^{\bar{j}}, ~~
 G_L^- = g_{i\bar{j}} \psi_L^{\bar{j}} \pa X^i \comma           \nono
	G_R^+ & = & g_{i\bar{j}} \psi_R^i  \bar{\pa} X^{\bar{j}}, ~~
G_R^- = g_{i\bar{j}} \psi_R^{\bar{j}} \bar{\pa} X^i\stop                
\eeqar
In addition, in order to preserve half of the spacetime supersymmetry,
we should take into account the spectral flow operator $\tofl$
defined by
\beq 
	\tofl = \omy_{i_1\ldots i_d} \psi^{i_1} \ldots \psi^{i_d} 
\comma
\eeq
Here $\Omega$  is the holomorphic $d$-form on the Calabi-Yau $d$-fold
and $J_L  = i \pa \phi_L$.  
Note that, in this convention, the $N=1$ supercurrent is
generated by
\beq \label {eq:n1_2}
	G = G_L^+ + G_L^-
\stop
\eeq

In order to represent a BPS saturated 
state in spacetime, the boundary must preserve half of the 
spacetime supersymmetry.  Thus we require
the boundary state to be invariant under a linear combination of 
the left and right 
$N=2$ algebra extended by the spectral flow operators.  
Consistency restricts the linear combination to correspond to the 
automorphism group of the algebra. The automorphism is
$O(2)$ for $N=2$ SCA and  $Z_2$ for $N=1$.
Since the supercurrent $G$ is gauged, its form should be preserved. 
Thus we are left with a $Z_2\times Z_2$-wise choice:

\noindent
{\bf  A-type boundary condition:}\footnote
	{In this section we write boundary conditions in
the notation appropriate for the open string 
	channel.}
 \beq \label {eq:n2a_bc}
	J_L = - J_R,~~~~   
	G_L^+ = \pm G_R^-,~~~~
	\tofl = e^{-i \phi_R} \stop
\eeq
\noindent
{\bf B-type boundary condition:}
\beq \label {eq:n2b_bc}
	J_L = + J_R,~~~~
	G_L^+ = \pm G_R^+,~~~~    
	\tofl = (\pm1)^d e^{i \theta} \tofr \stop 
\eeq
The phase factor $e^{i\theta}$ will be determined later.
In the A-type boundary condition, it can be absorbed in the definition of 
$\Omega$.  This is why we did not put the phase factor in
(\ref{eq:n2a_bc}). Clearly both  A-type and  B-type boundary conditions
preserve the $N=1$ SCA
\beq \label {eq:n1_bc}
	T_L = T_R,~~~~~
	G_L = \pm G_R \comma
\eeq
where $T$ denotes the stress tensor.
It should be noted that the mirror symmetry exchanges the A-type and the
B-type boundary conditions. 


\subsection {$N=4$ SCFT}
In the case of string 
compactification 
on K3, the spectral flow operators have the conformal weight $1$.
Combined with the $U(1)$ current $J$, they form the affine $SU(2)$ 
algebra and $N=2$ SCA is extended to $N=4$. 
For later convenience, let us write the holomorphic 2-form 
and the K\"ahler form as
\beq \label {eq:k3_k}
	\omy = k^1 + i k^2,~~~~~
	  k = k^3\stop
\eeq
The $SU(2)$ currents are then
\beqar \label {eq:n4j_d}
	J^I = k^I_{\mu \nu} \psi_L^{\mu} \psi_L^{\nu}~~~~(I=1,2,3)\comma
\eeqar
where the indices $\mu, \nu$ refer to real coordinates on K3.

In addition to $G^\pm$, we have two more supercurrents, which together
 with the original two form a \ul{4} of $SO(4)$, the automorphism group of $N=4$ SCA.
The automorphism consists of the internal and the external parts,
$SU(2)_c\times SU(2)_f$, where $SU(2)_c$ is generated by the $SU(2)$
currents $J^a$ and $SU(2)_f$ is the external automorphism of the 
N=4 SCA \cite{BV}. We can then organize the
four supercurrents as  $(\ul{2}, \ul{2})$
of $SU(2)_c\times SU(2)_f$ as 
\beqar \label {eq:n4g_d_22}
	G^{+-} &=& g_{i\bar{j}} \psi_L^i \pa X^{\bar{j}}, ~~
G^{++} = \omy_{ij} \psi^i_L \pa X^j, \nono
	G^{-+} &=& g_{i\bar{j}} \psi_L^{\bar{j}} \pa X^i, ~~
 G^{--} = \bar{\omy}_{\bar{i} \bar{j}} \psi_L^{\bar{i}} 
		\pa X^{\bar{j}}\stop
	\label {eq:n4g_1_22}
\eeqar

In this notation, the $N=1$ supercurrent $G$ is
\beq
	G = G^{+-} + G^{-+} \comma
\eeq
which is a singlet under the diagonal action of
$SU(2)_c \times SU(2)_f$. Since $G$ is fixed, 
a general boundary condition which
preserves both the N=4 and N=1 should only involve
the diagonal subgroup of $SU(2)_c \times SU(2)_f$,
i.e. $SO(3)$ in the full automorphism $SO(4)$.
By decomposing the four supercurrents into
$\ul{3}$ and $\ul{1}$ of $SO(3)$, the most general
boundary condition is written as
\beq \label {eq:n4_bc_4}
	J_L^I = U^I_{~J} J_R^J,~~~~
	G_L^I = \pm U^I_{~J} G_R^J,~~~~
	G_L = \pm G_R, ~~~(I,J=1,2,3)
	\comma
\eeq
where $U \in SO(3)$. 

\subsection {Geometric realization -- general case}

We would like to find out how the above classification of
supersymmetric boundary conditions corresponds to that of D-branes
in a Calabi-Yau manifold $M$. 
In this section, we seek this identification
in the large volume limit of $M$, where
we can treat the sigma-model semi-classically.  

We begin by noting that \qeq {n1_bc} is solved by 
\beq \label {n1_soln}
	\pa X^{\mu} = R^{\mu}_{~\nu} \bar{\pa} X^{\nu},~~~~~       
\psi_L^{\mu} = \pm R_{~\nu}^{\mu} \psi_R^{\nu} \stop
\eeq
for some matrix $R$ provided it satisfies 
\beq
   g_{\mu\nu} R^\mu_{~\rho} R^{\nu}_{~\sigma} = 
  g_{\rho\sigma} \stop
\eeq 
The eigen-vector of R with eigen-value $(-1)$ gives
the Dirichlet boundary condition for $X$, and thus
should correspond to directions normal to the D-brane.
If the matrix $R$ is symmetric, the
orthogonal directions are also eigenvectors of $R$
with eigen-values $(+1)$, and thus they obey the Neumann boundary
condition corresponding to the tangential directions
to the D-brane. In general, however, $R$ does not have to
be symmetric, and this gives rise to a mixed Neumann-Dirichlet condition. 
As we will see, this corresponds to the case when the $U(1)$ gauge
field on the D-brane worldvolume has non-zero field strength. 

In the neighborhood of a $p$-cycle $\gamma$ on the Calabi-Yau $d$-fold,
we can choose local coordinates such that $x^A$ ($A=1,...,p$) are
coordinates on the cycle and $y^a$ ($a=1,...,2d-p$) are for the
directions normal to $\gamma$. Clearly $(2d-p)$ is equal to the number
of $(-1)$ eigen-values of $R$.

Suppose the D-brane wrapping on
$\gamma$ gives the B-type boundary condition. 
	It follows from \qeq{n2b_bc} that $R$ should satisfy
\beqar 
	k_{\mu\nu} R^{\mu}_{~\rho} R^{\nu}_{~\sigma} &=& k_{\rho\sigma}\comma        
	\nonumber\\
	\omy_{\mu_1\ldots\mu_d} R^{\mu_1}_{~\nu_1}\ldots R^{\mu_d}_{~\nu_d} 
	&=& e^{i \theta} \omy_{\nu_1\ldots\nu_d} \stop
\label{anoy}
\eeqar
The first of these equations implies
\beq \label {n2bk_geo_more}
	k_{Ab} = 0 \comma
\eeq
namely the K\"ahler form $k$ must be block diagonal on $\gamma$ in
the tangential and the normal directions to $\gamma$. 
Since $k$ is nondegenerate, $k_{A B}$ and $k_{ab}$ must also be 
nondegenerate.  This means the dimensions $p$ of the cycle must be
even.  Because $k$ is block diagonal, we can use it to define
almost complex structure on the cycle.  In fact it is integrable and
defines a complex structure on the cycle.  Thus $\gamma$ is a holomorphic
submanifold of $M$.  In the complex coordinates, the 
nonvanishing components of the top form $\Omega$ has $p/2$ holomorphic indices 
tangential to $\gamma$ and $d - p/2$ holomorphic indices normal to it.  
This determines the phase $e^{i\theta}$ in (\ref{anoy}) 
in terms of the background gauge 
field on $\gamma$.  In particular when the gauge field is flat, we find
$e^{i\theta} = (-1)^{d-p/2}$.

On the other hand, if the cycle corresponds to the A-type boundary condition,
\qeq{n2a_bc} implies
\beqar \label {n2ak_geo}
	k_{\mu\nu} R^{\mu}_{~\rho} R^{\nu}_{~\sigma} &=& -
k_{\rho\sigma}\comma\nonumber\\
	\omy_{\mu_1\ldots\mu_d} R^{\mu_1}_{~\nu_1}\ldots R^{\mu_d}_{~\nu_d} 
	&=&  \bar{\omy}_{\nu_1\ldots\nu_d}  \stop
\eeqar
If the background gauge field on $\gamma$ is flat, 
$R$ squares to the identity matrix.  In this case, the
first of the above equations implies  
\beqar 
k_{ab} = 0, ~~~~k_{AB}=0 \stop
\eeqar
Since $k$ is nondegenerate, this is possible only if $p = d$.
Thus a cycle without a gauge field must be middle-dimensional. 
In this case, all the components of the holomorphic $d$-form
$\omy$ are related to $\omy_{A_1 \cdots A_d}$ as
\beq 
\omy_{a_1\cdots a_m A_{m+1} \cdots A_d} 
  \sim k^{~A_1}_{a_1} \cdots k^{~A_m}_{a_m}  \omy_{A_1 \cdots A_d} \comma
\eeq
for $m=1,...,d$.
Since $\omy\wedge\bar{\omy}$ is proportional to the volume form of the
$d$-fold, 
it follows that the 
pull-back of $\omy$ onto the cycle is proportional to its volume form.
We note that the same geometric condition for supersymmetric cycles
also arises from the low-energy effective worldvolume action 
of the supermembrane \cite{BBS} in the case of $p=3$.
It is easy to generalize this to the case with background gauge field. One can
see that (\ref{n2ak_geo}) implies $p=d, d+2, ..., 2d$. The reason for
this will become clear in the later sections. 

\subsection {Geometric realization - K3 case}
	In the case of K3, \qeq{n4_bc_4} states that
$k^I$ ($I=1,2,3$) behave as 
\beq \label {eq:n4_geo}
	k^I_{\mu\nu} R^{\mu}_{~\rho} R^{\nu}_{~\sigma} = U^I_{~J}
k^J_{\rho\sigma} \stop
\eeq
on the cycle $\gamma$. 
By going through some linear algebra, we find that the conjugacy class of the 
rotation $U$ is completely determined by the gauge field.  For example, in the absence
of the gauge field, the matrix $U$ is equal to $1$ for 0-cycle and 4-cycle 
while it is in the conjugacy class of $\pi$-rotation for 2-cycle.  
To understand this more 
geometrically, we diagonalize $U$ as
\beq \label {eq:k3_diag_u}
	U = M^t \left(    \begin{array} {rrr} 
					\cos \theta             &-\sin \theta   & 0  \\
					\sin \theta             &\cos \theta    & 0  \\
					0
& 0     &       1
				\end{array} \right)
			M
\stop
\eeq
By introducing a new basis by $M \in SO(3)$ rotation
\beq
  \tilde{k}^I = M^I_{~J} k^J \comma
\eeq
\qeq{n4_geo} is expressed as
\beqar  \label {eq:k3_di_bas}
	\tilde{k}^3_{\mu\nu} R^{\mu}_{~\rho} R^{\nu}_{~\sigma} &=&
\tilde{k}_{\rho\sigma}\comma  \nono
	\tilde{k}^\pm_{\mu\nu} R^{\mu}_{~\rho} R^{\mu}_{~\sigma} 
			&=& e^{\pm i \theta} \tilde{k}^\pm_{\rho\sigma}
\stop
\eeqar
Comparing this with the analysis of the B-type boundary condition 
in the previous subsection, we see that the cycle $\gamma$ is
a holomorphic submanifold of K3 with respect to the complex structure such that
$\tilde{k}^3$ is a K\"ahler form and $\tilde{k}^+$ is a holomorphic
2-form. Namely the $SO(3)$ rotation by $U$ reflects the $SO(3)$-wise
choice of complex structure for a given metric on K3. This result also
agrees with the analysis in \cite{BBS}, \cite{bsv2}.

\subsection {Summary}

	We now summarize our classification of boundary conditions.  
For each complex dimension $d$ of the Calabi-Yau manifold, we designate 
allowed values of $p$ (real dimensions of the cycle) and their possible 
boundary conditions by type A, B or the one parameterized by $SO(3)$.
\begin {center}
\begin {tabular}  {||l|r|r|r|r|r|r|r|r|r|r|r|r|r|r|r|r||} \hline
$d$     &1      &       &       &2      &       & 
 &  3   &       &       &       &       
		&4      &       &       &       &       \\ \hline
$p$     &0      &1      &2      &0      &2      &4      &0      &2      &3      &4      &6      
		&0      &2      &4      &6      &8      \\ \hline
Condition
	&B      &A      &B      &B  &$SO(3)$  &B  &B      &B      &A      &B      &B      
		&B      &B      &A/B &B &B  \\ \hline
\end {tabular}
\end {center}
This table is for the case with flat gauge field on $\gamma$. It is straightforward
to generalize this to the case with non-zero gauge field strength.    

One may notice that $p=3$ and $5$ for $d=4$ are not included in the
table\footnote{We would like to thank C. ~Vafa for drawing our
attention to this.}, even though there are Calabi-Yau 4-folds 
with non-trivial $H_3$. From the above analysis it is clear
that, provided the sigma-model for the 4-fold has only one set of $N=2$ SCA, 
one cannot construct a boundary condition at the SCFT level
corresponding to a 3-cycle which preserves half of the spacetime
supersymmetry. One arrives at the same conclusion by extending 
the analysis of \cite{BBS} to the case when the membrane wraps 
around a 3-cycle in a 4-fold. On a generic 4-fold
with $SU(4)$ holonomy, there are two covariantly constant spinors 
$\epsilon_1$ and $\epsilon_2$ of the same chirality. One then finds
that no linear combination of $\epsilon_1$ and $\epsilon_2$ can
generate spacetime supersymmetry which preserves the membrane
configuration. This does not mean that there is no Calabi-Yau 4-fold with a
supersymmetric 3-cycle. To the contrary, one can construct orbifold
examples which have such cycles. In these examples, however, the $N=2$
SCA is extended and thus the above classification is not applicable. 
Thus, an existence of a supersymmetric 3-cycle should imply
an extension of the worldsheet $N=2$ SCA. 
In general, a 4-fold can have a  holonomy group
$Spin(7)$, $SU(4)$, $Sp(2)$ or $SU(2)\times SU(2)$
\cite{Joyce}. The last two cases correspond,
 for instance, to the manifolds
$T^4\times K3$ and $K3\times K3$ respectively, and the associated worldsheet
algebras are extensions of the 
$N=2$ SCA. The generalization of the above classification of 
boundary conditions to these cases is straightforward.
The analysis for the $Spin(7)$ holonomy case will be reported
elsewhere \cite{OOZ}.
 
\section{Boundary states for D-branes}

In this section, we examine the properties of the boundary states
for D-branes wrapping on the supersymmetric cycles discussed
in the previous section. We will show how the
geometric data of the cycles are encoded in the boundary states.

\subsection{Supersymmetric boundary states}

Given the Virasoro algebra or its extension,
there is a definite procedure for constructing 
a conformally invariant 
boundary state, where the left and right generators of the 
algebra are linearly related,
starting from each highest weight state of the algebra. 
Denote by $|j,n\rangle,~ \overline{|j,n\rangle}$ orthonormal basis of the 
representations $j$ of
the holomorphic and antiholomorphic algebras respectively. 
It has been shown by Ishibashi \cite{Ishibashi} that 
\beq
|j\rangle\rangle = \sum_n |j,n\rangle\otimes U\overline{|j,n\rangle}
\comma
\label{ishi}
\eeq
is such a state, where $U$ is an anti--unitary matrix which
preserves the highest weight state $|j \rangle$.
A boundary state is in general a linear combination of
$|j \rangle \rangle$.

Type II strings compactified on Calabi-Yau spaces 
posses the worldsheet $N=2$ SCA in both the
left and right sectors. As we saw in the previous section,
a D-brane wrapping on a supersymmetric cycle preserves 
a linear combination of the left and right $N=2$ algebras. 
We would like to study the correspondence,
D-branes $\leftrightarrow$ boundary states, 
for D-branes wrapped on supersymmetric cycles
in Calabi-Yau spaces. In particular, given a D-brane, we would like 
to find the highest weight states
that appear in its boundary state and their multiplicity,
and conversely for a given boundary state we would like to find 
the D-brane configuration.

Recall from the analysis of section 2 that, for the closed strings,
there are two types of supersymmetric boundary conditions:
For middle-dimensional cycles, we have
\beq
G^+_L = \pm i G^-_R,~~~~~ 
G^-_L = \pm i G^+_R,~~~~~
J_L = J_R 
\comma
\eeq
and for even-dimensional cycles
\beq
G^+_L = \pm i G^+_R,~~~~~ 
G^-_L = \pm i G^-_R,~~~~~ 
J_L = -J_R 
\stop
\eeq
Here we are using the notation appropriate for the closed string
channel\footnote{$J_R \rightarrow - J_R$ and $G_R^\pm \rightarrow i G_R^\pm$
compared to the notation in section 2.}.  
They are called the A-type and the B-type boundary conditions.
For the K3 case, the boundary conditions are parameterized by $SO(3)$
corresponding to the $SO(3)$-wise choice of complex structures for a given metric
on K3. The boundary states realizing the A and B-type conditions should then 
satisfy
\beq
(G^+_L  \mp iG^-_R )|B\rangle = 0, ~~ 
(G^-_L  \mp iG^+_R)|B\rangle = 0,~~
(J_L - J_R) |B\rangle = 0 
\comma
\label{abc}
\eeq
or 
\beq
(G^+_L \mp iG^+_R)|B\rangle = 0,~~
(G^-_L \mp iG^-_R)|B\rangle = 0,~~ 
(J_L + J_R)  |B\rangle = 0
\comma
\label{bbc}
\eeq
depending on whether the boundary conditions are
A-type or B-type. Let us examine the properties of
these boundary states.

\subsection{A-type boundary condition}

Let us consider first the A-type boundary condition
corresponding to middle- dimensional cycles.
The boundary state can be expanded in terms of the Ishibashi states as
\beq
|B\rangle = \sum_{a} c^{a} |a\rangle\rangle 
\comma
\label{BS}
\eeq
where the sum is over the highest weight states
of the $N=2$ algebra which appear in the Hilbert space
of the sigma-model for the Calabi-Yau space $M$. 
They may be chiral primary states or non-chirals. 
According to our convention (\ref{eq:n2g_d}), 
complex moduli of $M$ are associated to
$(c,c)$ and $(a,a)$ primary states and
K\"ahler moduli are included in 
and $(c,a)$ and $(a,c)$.

The requirement that $(J_L-J_R)=0$ at the boundary implies
$q_L=q_R$ for the $U(1)$ charges  and thus a selection rule for the
conformal fields that can contribute to the boundary state.
In particular, this means that
the coefficients in front of the $(c,a)$ and $(a,c)$ primaries
are zero. In the following we will find an explicit form 
for the coefficients $c^a$ for the $(c,c)$ and $(a,a)$ 
chiral primary states.

For the sigma-model, the $(c,c)$ primaries with charge $(q,q)$
correspond to elements of the middle cohomology
$H^{q,d-q}(M)$ where $d = {\rm dim}_C M$. 
It is straightforward to show that the coefficient $c^{a}$ 
corresponding to the $(c,c)$ primary state is given by
\beq
c^{a} = \eta^{ab} \langle 0_{top} | \phi_b (z,\bar{z}) |B\rangle_{Ramond-Ramond}
\comma
\label{bcoef}
\eeq
where
$\langle 0_{top} |$ is the topological vacuum of the A-model,
$\eta^{ab}$ is the topological metric, and
$\phi_b$ is the $(c,a)$ primary field
associated to $\omega_b \in H^{q,d-q}(M)$. 
By the A-model, we mean the one with the topological twist
such that $G_L^+$ and $G_R^-$ become one-forms on the 
worldsheet\footnote{Thus the topological vacuum
$\langle 0_{top}|$
has charges $(-d/2, +d/2)$. Since the $(c,a)$ primary field $\phi_b$
carries charges $(q, q-d)$, the total charges of 
$\langle 0_{top} | \phi_b$ is $(q-d/2, q-d/2)$ satisfying the
selection rule.}. Since $\phi_b$
is physical in the A-model, and one may regard $c_a = \eta_{ab} c^b$ as a topological 
string amplitude on a disk with a puncture at $z$.  

The coefficient $c_a$ may in principle depends on
the K\"ahler moduli $(t^i, \bar{t}^i)$ ($i = 1,..., h^{1,1}$)
as well as the complex moduli of $M$. To compute
$\partial_{\bar{t}}$ of $c_a$, we insert $G^+_L G^-_R \bar{\varphi}_i$
onto the disk, where $\bar{\varphi}_i$ is an $(a,c)$ primary field
with $(q_L,q_R)=(-1,1)$. Since both $G_L^+$ and $G^-_R$ are one-forms
in the B-model, we can employ the standard contour deformation
argument in the topological field theory. Taking into account
the boundary condition $G_L^+ = \pm i G_R^-$, one finds that
the result of this insertion is zero. Thus $c_a$ is holomorphic
in $t^i$ and therefore the instanton approximation to $c_a$ is
exact. 

Furthermore one can also show that $c_a$ is independent
of $t^i$. One way to show this is to do the instanton expansion
explicitly and verify that the instanton correction vanishes due to the
fermion zero modes. 

Another way to show this is to insert $G^-_L G^+_R \varphi_i$
where $\varphi_i$ is a $(c,a)$ primary field with $(q_L, q_R)= (-1,1)$.
In this case, both $G^-_L$ and $G^+_R$ are two-forms on the disk
and we cannot immediately deform their contours.
On the disk with one puncture at $z$, there
is a global holomorphic $(-1)$ form $\xi(w) = (w-z)(w-\bar{z})$. 
By multiplying $\xi$,
we can convert $G^-_L$ into one-form and we can use the contour
deformation argument. Since $\xi(w)$ vanishes at $w=z$, where
$\phi_a$ is located, we can move the contour to the Dirichlet
boundary where we can convert $\xi G_L^-$ into 
$\bar{\xi} G_R^+$ since $\xi$ is real-valued on the boundary
(We chose the boundary to be ${\sl Im}~ w = 0$.). We can them
move $\bar{\xi} G_R^+$ back and the contour slips out of the
disk. Thus we have shown that $\partial_{t^i}$ of $c_a$
also vanishes. This reasoning is similar to the one which shows
that the topological metric of the A-model does not receive
the instanton correction. 

Since $c_a$ is independent of the K\"ahler moduli, we can take
the large volume limit in (\ref{bcoef}) to show
\beq
c_{a}(\gamma) = \int_\gamma \omega_a
\comma
\label{classicalbcoef}
\eeq
where $\gamma$ is the supersymmetric cycle in question. 
Thus  the chiral primary part of
the boundary state is determined entirely by the homology
class of the cycle $\gamma$.

This in particular means that 
the chiral primary part 
\beq
 | \gamma \rangle = \sum_{\phi_a: (c,c)} c^a | a \rangle_{Ramond-Ramond}
\comma
\eeq
of the boundary state is a flat section of the so-called improved
connection \cite{Strominger}, \cite{Candelas}, \cite{CV} 
for the bundle of Ramond vacua
over the moduli space of $N=2$ superconformal field theories
(for a review, see also section 2 of \cite{BCOV}).
Since it plays an important role in the case of
the B-type boundary condition in the following, let us demonstrate 
this fact explicitly here. Let us organize 
the basis of $H^d(M)$ as $\omega_{0} \in H^{d,0}$,
$\omega_\alpha \in H^{d-1,1}$ ($\alpha = 1,..., h^{d-1,1}$), etc. 
Then we find
\beq
\frac{\partial  c_0}{\partial \bar{y}^\alpha}  = 0,~~
\frac{Dc_0}{D y^\alpha}  = c_\alpha,~~{\rm etc,}
\eeq
where $y^\alpha$ are the complex moduli of $M$ and
$D$ is the covariant derivative on the vacuum line
bundle ${\cal L}$ 
over the moduli space of the $N=2$ theories. 
These equations can be summarized as
\beq
   \nabla_\alpha | \gamma \rangle = 0, ~~
   \bar{\nabla}_{\bar{\alpha}} | \gamma \rangle = 0
\comma
\eeq
where
\beq
   \nabla_\alpha = D_\alpha - C_\alpha,
~~ \bar{\nabla}_{\bar{\alpha}} = \bar{D}_{\bar{\alpha}} 
     - \bar{C}_{\bar{\alpha}}
\comma
\eeq
and $C_\alpha$ is the multiplication by the Yukawa
coupling. 

This in particular means that 
$c_a$ for $\omega_a \in H^{d-1,1}$ etc, is obtained by
acting  with $D_\alpha$ on $c_0$. Thus the chiral primary 
part of the coefficients in (\ref{BS}) is completely determined 
by computing the period 
\beq
  c_0(\gamma) = \int_\gamma \Omega
\comma
\label{period}
\eeq
of the holomorphic $(d,0)$-form. 
To be precise, this is the case  when the complex dimension
of the Calabi-Yau manifold is less than 4.
When $d \geq 4$, there is some subtlety since there may be
an element $\omega_a$ of $H^{d-q,q}$ with $q \geq 2$ 
which is not generated by differentiating
$H^{d,0}$ with respect to the complex moduli. If that is a case,  
we have to evaluate (\ref{classicalbcoef}) for such $\omega_a$
separately. Understanding
how this procedure works for $d \geq 4$ would help clarify issues
on the mirror symmetry in higher dimensions \cite{GMP}.

\subsection{B-type boundary condition}

For an even-dimensional cycle $\tilde{\gamma}$, the boundary states satisfy
the B-type condition $(J_L + J_R) | \tilde{B} \rangle = 0$. 
Thus the coefficients $\tilde{c}^a$ for the expansion
\beq
   | \tilde{B} \rangle = \sum_a \tilde{c}^a |a\rangle \rangle
\comma
\eeq 
vanish for the $(c,c)$ and $(a,a)$ primary states. On the other
hand, the coefficients
for the $(c,a)$ primaries are obtained by
\beq
  \tilde{c}^a = \tilde{\eta}^{ab} \langle \tilde{0}_{top}
  | \tilde{\phi}_b(z,\bar{z}) |\tilde{B} 
   \rangle_{Ramond-Ramond}
\comma
\eeq
where $\langle \tilde{0}_{top} |$ is the topological vacuum of the
B-model, $\tilde{\eta}^{ab}$ is the topological metric 
and $\tilde{\phi}_a(z,\bar{z})$ is the $(c,c)$ primary field associated
to $\tilde{\omega}_a$ in the vertical series of the cohomologies
$H^{vertical}(M) = \oplus_{q=0}^d H^{q,q}(M)$.
The B-model is defined in such a way that $G_L^+$ and $G_R^+$ behave
as one-forms\footnote{Thus the topological vacuum 
$\langle \tilde{0}_{top} |$ has charges $(-d/2, -d/2)$
while $\tilde{\omega}_a$ carries $(d-q,q)$. Combined, they
satisfy $q_L = - q_R$ as required. }.

By repeating the contour deformation argument as in the case of the
A-type boundary condition, one finds that $c^a$ is independent of
the complex moduli $y$, but may depend on the K\"ahler moduli $(t,
\bar{t})$. We now present two arguments to show 
that the $(c,a)$ primary part of the boundary state
\beq
 | \tilde{\gamma} \rangle =  \sum_{\tilde{\phi}_a: (c,a)} \tilde{c}^a 
| a \rangle_{Ramond-Ramond}
\comma
\eeq
is ``flat'' with respect to the improved connection over the
K\"ahler moduli space. This determines the $(t, \bar{t})$
dependence of $\tilde{c}^a$.

A simple way to show this is to use the mirror symmetry. Since the mirror
symmetry transforms the A-type boundary condition into the B-type,
the flatness property of the state $| \gamma \rangle$ over the complex
moduli space for the middle-dimensional
cycle $\gamma$ should imply the flatness of $| \tilde{\gamma}
\rangle$ over the K\"ahler moduli space 
for the even-dimensional cycle $\tilde{\gamma}$ provided 
$\gamma$ and $\tilde{\gamma}$ are related to each other by the
mirror transform.

In the next section, we will use the 
flatness of $| \tilde{\gamma} \rangle$
to study the mirror symmetry between the D-branes. 
For the sake of completeness, we therefore give another argument
for the flatness which stands independently of the mirror symmetry. 
 To take a derivative
of $\tilde{c}_a$ with respect to the K\"ahler moduli $t^i$, we insert
$G_L^- G_R^+  \varphi_i $ on the disk, where $\phi_i$ is 
a $(c,a)$ primary field corresponding to an element of $H^{1,1}$. 
Unlike the case of the complex moduli derivative, however,
this does not yet give us $D_i \tilde{c}_a$ since $G_L^- G_R^+
\varphi_i $ is divergent at the Dirichlet boundary. The covariant
derivative $D_i$ must be defined in such a way that the
contribution from the boundary is removed. 
Since $G^+_R$ is a one-form in the $B$-model,
we can deform its contour on the disk. By taking into account the
boundary condition (\ref{bbc}), one finds that $G_L^- G_R^+
\varphi_i$ becomes $\partial \varphi_i$. The integral of 
$\partial \varphi_i$ over the disk with the puncture reduces
to two surface integrals, one around the puncture at $z$ and
another around the Dirichlet boundary. The former can be
evaluated using the Yukawa coupling since it is related to
the OPE of $H^{1,1}$ and $H^{q,q}$. The latter is
canceled by the covariantization. This shows
\beq
  (D_i - C_i ) | \tilde{\gamma} \rangle = 0 
\comma
\eeq
and similarly 
\beq
  (\bar{D}_{\bar{i}} - \bar{C}_{\bar{i}} ) | \tilde{\gamma} \rangle = 0 
\stop
\eeq

The flatness of $| \tilde{\gamma} \rangle$ implies that
the coefficient $\tilde{c}_0$ corresponding to the top cohomology
$H^{d,d}$ is holomorphic with respect to the K\"ahler moduli.
 It also implies that
the rest of $\tilde{c}_a$ is obtained by taking derivatives of
$\tilde{c}_0$ with
respect to $t$.  Since $\tilde{c}_0$ is holomorphic in $t$, the instanton 
approximation is exact, i.e. $\tilde{c}_0$ can be expressed as
a sum over holomorphic maps from the disk to 
$M$ such that the boundary of the disc is mapped to the cycle
$\tilde{\gamma}$. 
When $\tilde{\gamma}$ is $2q$-dimensional, the contribution
from the constant map can be evaluated by taking the large
volume limit as
\beq 
  \tilde{c}_0(\tilde{\gamma}) \sim
 \int_{\tilde{\gamma}} k^q + O(e^{2\pi i t})
\comma
\label{largevol}
\eeq
where $k = \sum_i t^i k_i$ and we choose $k_i$ to be the basis of 
$H^{1,1}(M;Z)$. 

The instanton corrections to $\tilde{c}_0$ are obtained by replacing
the classical intersections in (\ref{largevol}) by quantum
ones in an appropriate sense. This in particular implies
that  $\tilde{c}_0$ for 0 or 2-cycle does not receive an instanton
correction since the image of the holomorphic map of the disc does not 
intersect with the homology dual to $k_i$ in these cycles. 
In the next section, we will find that this in fact
is consistent with the mirror symmetry.

The expressions (\ref{largevol}) in particular means that
the large volume limit of $\tilde{c}_0$ is a homogeneous polynomial of $t$
and the dimensions of the cycle $\tilde{\gamma}$ is characterized by the
degree of the polynomial. One may be worried  
that this statement is not invariant under the integral shift 
of the theta parameters of the sigma-model, $t^i \rightarrow t^i  + m^i$
($m^i \in Z)$. In fact this shift should mix cycles of different
dimensions. Consider a cycle $\tilde{\gamma} \in H_{vertical}(M;Z)$ 
and decompose it as 
\beq
\tilde{\gamma} = \sum_{q=0}^d \tilde{\gamma}_q
\comma
\eeq
where $\tilde{\gamma}_q \in H_{q,q}(M;Z)$. The equation (\ref{largevol})
can then be rewritten as
\beqar
  \tilde{c}_0(\tilde{\gamma})
    &=& \sum_q \int_M k^q \wedge  \tilde{\gamma}_q^* \nonumber
    \\
   &=& \int_M e^k \wedge \left( \sum_q q! \tilde{\gamma}_q^* \right) 
\comma
\eeqar
where $\tilde{\gamma}_q^* \in H^{d-q,d-q}(M;Z)$ is the
Poincare dual of $\tilde{\gamma}_q$.  One then finds that the shift 
$k \rightarrow k + \omega$ with $\omega \in H^{2}(M;Z)$ mixes
$\tilde{\gamma}_q$'s as
\beq
  \tilde{\gamma}_q^* \rightarrow \sum_{n} {}_{q+n} C_n~
                      \omega^n \wedge \tilde{\gamma}_{q+n}^*
\stop
\label{monod}
\eeq
As we will see in the next section, this mixing
is in accord with the mirror symmetry. 

\section{Mirror symmetry}

The mirror symmetry transforms type IIA string on a Calabi-Yau
3-fold $M$ into type IIB string on the mirror $\widetilde{M}$. Since
type IIA string has even-dimensional D-branes while type IIB has
odd-dimensional ones, we expect that the mirror symmetry to
transform middle ($= 3$) dimensional cycles on $M$ into even-dimensional
cycle on $\widetilde{M}$. From the point of view of SCFT, 
the mirror symmetry transforms the A-type boundary condition (\ref{abc})
for the 3-cycle to the B-type boundary condition (\ref{bbc})
for the even-dimensional cycle. In this section, we will examine 
how this transformation between the supersymmetric cycles takes place.

The mirror symmetry is an isomorphism between the Hilbert spaces 
of the sigma-models on $M$ and $\widetilde{M}$ \cite{GP}. Thus if the cycles
$\gamma$ and $\tilde{\gamma}$ are related to each other by the mirror
symmetry, the corresponding boundary states $| B \rangle$
and $| \tilde{B} \rangle$ should be identified by the
isomorphism\footnote{To be precise, the boundary state $| B \rangle$
does not belong to the Hilbert space since it is not normalizable.
This problem can be easily avoided by considering 
 $q^{L_0} {\bar{q}}^{\bar{L}_0} | B \rangle$
for $|q| < 1$, for example.}. 

\subsection{Mirror map between cycles}

Suppose the boundary state $| B \rangle$ for a 3-dimensional cycle
$\gamma$ in $M$ is mapped to the boundary state $| \tilde{B} \rangle$
for  an even-dimensional cycle $\tilde{\gamma}$ in $\widetilde{M}$ 
under the mirror transformation. Since the chiral primary part
of the boundary states are characterized by $c_0$ and $\tilde{c}_0$
given in the previous section, they should be related to each other
under the mirror map.  For the 3-cycle $\gamma$, $c_0$ is given by
\beq
  c_0(\gamma) = \int_\gamma \Omega
\stop
\label{ac0}
\eeq
Since we know the large volume limit of $\tilde{c}_0$ as in 
(\ref{largevol}), we should compare it with $c_0$ in the corresponding
limit, which is called the large complex structure
limit \cite{Morrison} of $M$. 

In this limit, $H^{0,3}(M)$ aligns with
the lattice of $H^3(M;Z)$ \cite{Candelas}, \cite{al}. 
Thus we have a filtration of $H^3(M;Z)$ in a form of
\beq
 H^{0,3} \subset H^{0,3} \oplus H^{1,2}
   \subset H^{0,3} \oplus H^{1,2} \oplus H^{2,1}
   \subset H^3(M;Z)
\comma
\eeq
called the monodromy weight filtration \cite{morrison3}.
Accordingly we can choose a symplectic basis 
$\{ \alpha_I, \beta^I \}_{I = 0,...,
h^{2,1}}$ for $H_3(M;Z)$, 
\beq
  \alpha_I \cap \alpha_J = 0, ~~ \beta^I \cap \beta^J = 0, ~~
  \alpha_I \cap \beta^J = \delta_I^J
\comma
\eeq
such that $\alpha_0$ is the unique cycle dual to $H^{0,3}$ and
$\{ \alpha_0 , ... , \alpha_{h^{2,1}} \}$ spans the dual
of $H^{0,3} \oplus H^{1,2}$. The cycle $\alpha_0$ may 
also be characterized by the fact that it is invariant under
the monodromy of $H_3(M;Z)$ at the large complex structure limit
\cite{Morrison1}, \cite{Morrison2}. Note, on the other hand,
$\alpha_i$ with $i = 1 ,...,h^{2,1}$ may be shifted by 
$\alpha_0$ under the monodromy transformation.

With this choice of the basis for $H_3$,
the flat coordinates of the complex moduli space are given by
\beq
     s^i = \frac{X^i}{X^0}~~~(i=1,...,h^{2,1}(M) = h^{1,1}(\widetilde{M}))
\comma
\eeq
where
\beq
    X^0 = \int_{\alpha_0} \Omega, ~~
    X^i = \int_{\alpha_i} \Omega
\stop
\eeq
In the large complex structure limit $s \rightarrow \infty$
 the Schmid  orbit theorem \cite{Schmid} yields
\beqar     
  c_0(\beta^0) =  \int_{\beta^0} \Omega &=&   
\frac{1}{3!} X^0 d_{ijk}s^is^j s^k + \cdots 
\comma
\nonumber \\
 c_0(\beta^i) =  \int_{\beta^i} \Omega &=& 
-\frac{1}{2!} X^0 d_{ijk} s^js^k + \cdots 
\comma
\label{largecom1}
\eeqar
where $d_{ijk}$ is the large
complex structure limit of the Yukawa coupling.
					   
In order to construct the mirror map, we choose the standard gauge of the
special geometry, 
\beq
 c_0(\alpha_0) = \int_{\alpha_0} \Omega = 1
\stop
\label{largecom2}
\eeq
In this gauge, the flat coordinates are
\beq
 c_0(\alpha_i) = \int_{\alpha_i} \Omega = s^i 
\stop
\label{largecom3}
\eeq
By the mirror map, we may also use it as 
the flat coordinates for the K\"ahler moduli 
space of $\widetilde{M}$. In the large
complex structure limit, this mirror symmetry maps
the Yukawa coupling $d_{ijk}$ in (\ref{largecom1}) to
\beq
  d_{ijk} = \int_{\widetilde{M}} k_i \wedge k_j \wedge k_k
\stop
\label{largecom4}
\eeq

By comparing large volume limit (\ref{largevol}) of
$\tilde{c}_0$ for even-dimensional cycles in $\widetilde{M}$
with the large complex structure limit
(\ref{largecom1}) -- (\ref{largecom4}) of $c_0$
for $\{ \alpha_I, \beta^I \}$, 
we can immediately see how the mirror symmetry transforms
a D-brane wrapping on a 3-cycle in $M$ to
a D-brane wrapping on an even-dimensional cycle 
in $\widetilde{M}$. In particular, the 3-cycle $\alpha_0$ 
dual to $H^{0,3}$ in $M$ is a mirror image of a 0-cycle in $\widetilde{M}$, 
and the 3-cycles $\alpha_i$ ($i=1,...,h^{1,2}$) 
correspond to 2-cycles in $\widetilde{M}$. 
Thus the mysterious correspondence between the integral structures
of $H^3(M)$ and $H^{vertical}(\widetilde{M})$ pointed out in 
\cite{al} is now understood as the mirror symmetry
between the D-brane configurations.

While this paper is being typed, we received a preprint \cite{SYZ} by
Strominger, Yau and Zaslow where it is argued that the 
mirror of a 0-cycle in $\widetilde{M}$ should be a {\it toroidal} 3-cycle
in $M$. Our analysis here shows a mirror of the
0-cycle should be the 3-cycle $\alpha_0$ dual to $H^{0,3}$ in the
large complex structure limit of $M$. In
the case of the quintic defined by, 
\beq
  p(x) = x_1^5 + x_2^5 + x_3^5 + x_4^5 + x_5^5 - 5 \psi x_1x_2x_3x_4x_5
  = 0
\comma
\eeq
such a 3-cycle is in fact known to be $T^3$ \cite{cogp}. 
In the large complex structure limit
$\psi \rightarrow \infty$, the holomorphic 3-form becomes
\beq
  \Omega = 
     5 \psi 
     \frac{x_5 dx_1 \wedge dx_2 \wedge dx_3}
   {\partial p / \partial x_4} 
     \rightarrow - \frac{dx_1 dx_2 dx_3}{x_1x_2x_3} 
\comma
\eeq
and the 3-cycle dual to $\bar{\Omega}$ is $T^3$ surrounding $x_1=x_2=x_3=0$.
It would be very interesting
to see whether this feature of $H^{0,3}$
is true for a general $M$ with a mirror partner.

So far we have only looked at the large
volume limit of $\widetilde{M}$  and the corresponding large complex structure
limit of $M$. Fortunately, since the state $| \tilde{\gamma} 
\rangle$ and $| \tilde{\gamma}\rangle$
are flat sections over the moduli spaces, 
their correspondence can be traced to interiors
on the moduli spaces following the mirror map. 
We will demonstrate this through examples in section 5. 
If we go around a non-trivial cycle over the
moduli space, we have to deal with the monodromy problem,
which we will discuss below. 

\subsection{Open string instantons}

For the A-type boundary condition, the classical formula
\beq
  c_0(\gamma) = \int_\gamma \Omega
\eeq
is exact. On the other hand, the formula (\ref{largevol})
for $\tilde{c}_0(\tilde{\gamma})$ for the
B-type boundary condition is corrected by open string worldsheet
instantons, i.e. holomorphic maps from a disk to $\widetilde{M}$
such that the boundary of the disk is mapped to the cycle
$\tilde{\gamma}$. The mirror symmetry suggests that 
such open string instanton effects are expressed in terms of
the closed string instantons on $\widetilde{M}$.  

The mirror symmetry gives another proof for the fact 
that the formula (\ref{largevol}) for $\tilde{c}_0$ does not receive
the instanton correction when the cycle $\tilde{\gamma}$ is 0 or
2-dimensional. This is because the corresponding
formulae (\ref{largecom2}) and (\ref{largecom3}) for
$\alpha_I$ $(I=0,...,h^{2,1})$ are, by definition, exact. 

On the other hand, $\tilde{c}_0$ for 4 or 6-cycle does receive
instanton corrections. In the mirror picture, the exact formulae
for $c_0(\gamma)$ for $\beta^I$ $(I=0,...,h^{2,1})$ can be written
in terms of the prepotential ${\cal F}$ for $M$ as
\beqar
   c_0(\beta^0) = 2 {\cal F} - s^i \frac{\partial}{\partial s^i} 
        {\cal F} , \nonumber\\
   c_0(\beta^i) = \frac{\partial}{\partial s^i} {\cal F} 
\comma
\label{exact}
\eeqar
where we are working in the $X^0=1$ gauge appropriate for the mirror
symmetry. In $\widetilde{M}$, the prepotential
is related to the sum over closed string instantons 
as\footnote{We are using the same coordinates $s^i$ for
both the complex moduli of $M$ and the K\"ahler moduli of $\widetilde{M}$
related to each other by the mirror symmetry.}
\beq
- \frac{\partial^3}{\partial s^i \partial s^j \partial s^k} {\cal F}
  = d_{ijk} + \sum_{n} N(n) n_i n_j n_k 
 \frac{ e^{2 \pi i n_i s^i} }{1 - e^{2 \pi i n_is^i}} 
\comma
\eeq
where $N(n)$ is the number of rational curves on $\widetilde{M}$ of
the type $n=\{n_1,...,n_{h^{1,1}(\widetilde{M})}\}$.
By integrating this, we find 
\beq
 {\cal F} = -\frac{1}{3!} d_{ijk} s^is^js^k + a
    - \sum_{n} \sum_{m=1}^\infty \frac{N(n)}{(2 \pi im)^3}
 e^{2 \pi i  m n_i s^i} 
\comma
\eeq
where $a$ is a constant, presumably related to the four-loop term
in the $\beta$-function of the sigma-model \cite{cogp}. 
Substituting this into (\ref{exact}), we can extract the
open string instanton corrections to $\tilde{c}_0$ and express them in terms of 
of the number of the closed string instantons $N(n)$.

This suggests a relation between the moduli spaces of open and 
closed string instantons and the corresponding intersection theories.
One way to find such a relation may be to regard a closed
string instanton intersecting a supersymmetric cycle 
as a pair of open string instantons glued on the cycle. 

\subsection{Integral structure and monodromy}

It has been observed that in the large radius 
limit, the mirror symmetry maps the integer valued
homology $H_3(M;Z)$ to $\oplus_q H_{2q}(\widetilde{M};Z)$
in such a way that the monodromy is preserved
\cite{al}, \cite{Morrison1}.
Based on this, it was 
conjectured by Aspinwall and Morrison \cite{AM,Morrison2}
that the Ramond-Ramond fields
on the Calabi-Yau 3-fold should have
periodicity under the discrete shift reflecting these integral
structures. This would guarantee that the mirror symmetry can
be extended to the Ramond-Ramond fields configurations.
This periodicity should be a consequence of the coupling of
the Ramond-Ramond field to the worldvolume of 
the D-brane. In fact the mirror map between the D-branes
we found in the above is consistent with this picture. 

By requiring that the monodromy be preserved,
Morrison also pointed out \cite{Morrison2}
that the shift of the NS-NS $B$-field by $H^2(M;Z)$
should cause a certain rearrangement of the integral 
structure of the Ramond-Ramond fields of even ranks. 
This is also consistent with the mixing of the even 
dimensional cycles we found in (\ref{monod}).

Although the mixing of the cycles is required by the mirror symmetry,
one can also explain it without invoking the mirror. 
For the sigma-model without a boundary, the shift of $B$-field by
$H^2(M;Z)$ is a discrete symmetry. However, in the presence of a boundary, the coupling of
the $B$-field to the string world-sheet is accompanied by the coupling 
of a $U(1)$ gauge field $A$ to the boundary \cite{Witten}. 
Since the gauge invariant field strength is ${\cal F} = F - B$
where $F = dA$, the shift $B \rightarrow B + \omega$ with 
$\omega \in H^2(M;Z)$ is compensated by $F \rightarrow F + \omega$.
This effectively mixes cycles of different dimensions as in
(\ref{monod}). Below we will demonstrate this explicitly 
through examples. 

\section{Examples}
In this section we will present several examples to illustrate the general
results of the previous sections.
We will show explicitly how
starting with a D-brane wrapped on a middle-dimensional supersymmetric
cycle, depending on the D-brane configuration and  T-duality or
mirror
transformation, we can obtain different dimensionalities
for the dual configuration with gauge fields background.

\subsection{T-duality on tori}

Let us start with a general discussion of the duality map for tori and orbifolds.
As we discussed in section 2, the condition for $N=1$ SCA yields
\beq
\pa X^{\mu} = R^{\mu}_{~\nu} \bar{\pa} X^{\nu}
\comma
\label{R}
\eeq
where $R$ is an orthogonal matrix.
The requirement for having  a geometrical interpretation of a D-brane
without gauge fields background 
is more restrictive and implies that 
$R$ has to be a symmetric matrix and squares to the identity matrix. 
In this case, its eigen-values are $(+1)$ or $(-1)$ corresponding to the
tangential and normal vectors to the D-brane respectively.
To preserve the $N=2$ SCA, $R$ should further obey
\beq
k_{\mu\nu} R^\mu_{~\rho} R^\nu_{~\sigma} = \pm k_{\rho\sigma}
\comma
\eeq
where $\pm$ refers to the A and B-type boundary conditions and thus
to middle and even-dimensional cycles.
 
T-duality transformation is realized by
\beq
\pa X^{\mu} \rightarrow \pa X^{\mu}
\comma
~~~~\bar{\pa} X^{\mu} \rightarrow T^{\mu}_{~\nu}\bar{\pa} X^{\nu}
\comma
\eeq
where $T$ is the symmetric matrix implementing the duality transformation and
$T^2=1$.
In order for this to induce the mirror transformation, the sign of $J_R$
should be reversed while $J_L$ remain invariant. This means
\beq
k_{\mu\nu} T^\mu_{~\rho} T^\nu_{~\sigma} = - k_{\rho\sigma} 
\stop
\label{tdualmirror}
\eeq 
Thus, starting with a D-brane configuration and performing T-duality transformation
we will end up with a configuration satisfying the boundary condition
\beq
\pa X^{\mu} = \tilde{R}^{\mu}_{~\nu} \bar{\pa} X^{\nu}
\comma
\eeq
where $\tilde{R} = RT$ is an orthogonal matrix.
If the  matrix $\tilde{R}$ is symmetric and thus squares to the
identity matrix, 
the boundary condition has geometrical realization as a D-brane
without the $U(1)$ gauge field. This occurs if and only if 
\beq
\[R ,T\]=0
\comma
\label{comm}
\eeq
namely T-duality transformation commutes with the original D-brane
configuration.

When (\ref{comm}) is not satisfied,
we get a mixing between the Neumann and Dirichlet boundary conditions of the type induced
by a background gauge field. Since $\tilde{R} = RT$ is orthogonal, by a
coordinate transformation, we can alway bring it into the standard form,
\beq
 \tilde{R} =
\pmatrix{-1_{(2d-p)\times(2d-p)} & 0 \cr 0 & \(\frac{1-F}{1+F}\)
_{p\times p} \cr}
\comma
\eeq
where for some $p$ and an anti-symmetric matrix $F$. This implies the
Dirichlet boundary condition for the first $(2d-p)$ directions, while
the boundary condition for the second $p$ directions is
\beq
   \partial X^\mu = \left( \frac{1-F}{1+F} \right)^\mu_{~\nu}
   \bar{\partial} X^\nu
\stop
\eeq
Therefore the matrix $\tilde{R}$ describes a $p$-cycle with a
background gauge field $F$.

Whether $F$ is zero or not,  
the mirror symmetry exchanges 
odd and even-dimensional cycles when $d = {\rm dim}_C M$ is odd. 
In this case, the condition (\ref{tdualmirror}) for T-duality 
to be the mirror symmetry implies
${\rm det}T =-1$. On the other hand, ${\rm det}R = -1$ for an odd
dimensional cycle since the rotation matrix $(\frac{1-F}{1+F})$
has determinant $(+1)$.
Thus $\tilde{R} = R T$ for its mirror obeys
${\rm det} \tilde{R} = {\rm det} R \cdot {\rm det} T = +1$, i.e.
the mirror of the odd dimensional cycle is even-dimensional. 
If $R$ and $T$ commute, $F=0$ in the original cycle implies
$F=0$ for its mirror. 

Let us construct now a simple example to illustrate the above.
Consider the torus $T^2$ with real coordinates $(x,y)$, 
and a D-brane configuration defined by the Pauli matrix
$R=\sigma_1$. The Neumann boundary condition is imposed on the 1-cycle defined by
the vector $(1,1)$, while the Dirichlet boundary condition is imposed on
the vector orthogonal to it.
Then the mirror transformation is generated by T-duality
transformation along the $x$ coordinate, i.e. $T=-\sigma_3$. 
Clearly this $T$ does not commute with $R$. In fact
$\tilde{R}=-i\sigma_2=F$, and this has no $(-1)$ eigen-value,
namely there are no Dirichlet boundary conditions.
The configuration we got is that of a 2-cycle with background gauge field $F$.  

It is instructive to consider this example from a different viewpoint.
In the limit of the large complex structure, $\tau \rightarrow i\infty$,
the cohomology $H^{0,1}$ generated by $d\bar{z} = dx + \bar{\tau} dy$
gets aligned with the lattice $H^1(T^2;Z)$ generated by $dx$ and $dy$. 
In this limit, the cycle $(1,0)$ becomes dual to $H^{0,1}$
and the mirror map transforms it to a 0-cycle, as expected.
On the other hand, either $(0,1)$ or $(1,1)$ can be combined with
$(1,0)$ to make the symplectic basis of $H^1(T^2;Z)$. Since
$(0,1)$ is mirror to a 2-cycle without a gauge field, one may
regard $(1,1) = (0,1) + (1,0)$ as mirror to the 2-cycle with a 0-cycle
on it.  Though the filtration $H^{0,1} \subset H^1(T^2;Z)$ makes sense only in the large
complex structure limit, the mirror map between the cycles holds
even for finite value of $\tau$. The reason for this can be traced
back to the fact that the chiral primary part of the boundary state 
$| \gamma \rangle$ is a flat section over the moduli space of
complex structure, as we explained in section 4. 

This picture is correct as far as the homology goes, but 
a sum of the straight lines, $(0,0) \rightarrow (1,0)$ and $
(1,0) \rightarrow (1,1)$, is not actually supersymmetric since 
the combined cycle is not minimal. The diagonal line $(0,0) \rightarrow
(1,1)$ is shorter and thus costs less energy. 
In the mirror picture, this means that the 2-cycle with the $U(1)$
gauge field should be regarded as a ground state of the 0-cycle on  
the 2-cycles.

This simple example illustrates the mixing of cycles (\ref{monod}).
The D-brane worldvolume action has terms of the form \cite{Douglas}
\beq
S= \int_{2-cycle} C_0 {\cal F} + C_2
\comma
\eeq
where $C_0$ and $C_2$ are the Ramond-Ramond fields and 
${\cal F} = F-B$. A shift of $B$ by $H^2(T^2;Z)$
then mixes $C_0$ and $C_2$ corresponding to the mixing
of cycles. In the mirror picture, the shift $B \rightarrow B+1$
becomes the modular transformation $\tau \rightarrow \tau + 1$.
This sends the cycle $(0,1)$ (the 2-cycle in the mirror) 
to $(1,1)$ (the 0-cycle on  the 2-cycle in the mirror). 
Thus the mixing of the cycle (\ref{monod}) is
natural from the point of view of the coupling of the D-brane
to the $B$ field \cite{Witten} as well as the mirror symmetry. 


\subsection{Calabi-Yau orbifold}

In this section we discuss an example of a mirror pair of Calabi-Yau
orbifolds.
In fact the phenomena is basically similar to the tori cases, 
with some technicality related to the
correct choice of a ground state.
As an explicit example we will consider the  mirror of the Calabi-Yau 
orbifold $(T^2)^3/(Z_2 \times Z_2)$ 
which is constructed by the inclusion of a discrete torsion \cite{VW}.
Let us first discuss the orbifold without a discrete torsion.
The Calabi-Yau orbifold $(T^2)^3/\Gamma$ where
$\Gamma = Z_2 \times Z_2$  is defined by 
$z_i \rightarrow (-1)^{\varepsilon_i} z_i,~~ i=1,2,3$
such that $\prod_i (-1)^{\varepsilon_i} =1$.
Supersymmetric 2 cycles can constructed by projecting a $T^2$ in $(T^2)^3$ with respect to $\Gamma$.
 Similarly, supersymmetric 4-cycles can be obtained by projecting a product of two $T^2$'s with respect
to $\Gamma$.
The even-dimensional supersymmetric cycles are interesting in this example since the
twisted Ramond ground states contribute to $H^{1,1}$ and $H^{2,2}$. 
Thus the latter can show up in their 
boundary states. 

Consider, for instance, a 2-cycle boundary state where Neumann boundary conditions are imposed on the $z_3$
coordinate and Dirichlet boundary conditions on $z_1,z_2$. 
Orbifold boundary states are simply constructed as a sum of 
contributions from the untwisted 
and twisted sectors
\beq
|B\rangle_{orbifold} = |B\rangle_{untwisted} + \sum_{twisted~ sectors}|B\rangle_{twisted}
\comma
\eeq
with an appropriate projection on invariant states.

\noindent
{\bf untwisted sector}: 

The boundary state takes the form
\beq
|B\rangle_{untwist} = {\rm exp} \left( -\sum_{n=1}^{\infty}\frac{1}{n}(\sum_{i=1}^2
\alpha^i_{L,-n}\alpha^i_{R,-n}  + 
\alpha^{*3}_{L, -n}\alpha^3_{R, -n}) + {\rm c.c} 
\right)|0\rangle
\comma
\eeq
and projection is not required since the boundary state is $\Gamma$-invariant.
The fermionic part works similarly.

\noindent
{\bf twisted sectors:} 

There exist three twisted sectors corresponding to the three $\Gamma$ group elements. 
Consider, for instance, the twisted
sector corresponding to the generator $\alpha$,
$\alpha (z_1,z_2,z_3) = (-z_1,-z_2,z_3)$, where the $\beta$ and $\gamma$
are defined by a permutation of the signs. This implies half integer
modding for the first two coordinates and integer modding for the third.
The other twisted sectors are simply permutations of that.

Let us consider now the inclusion of a discrete torsion.
This simply amounts to a change in the projection operators in the twisted sectors.
Thus in the sector twisted by $\alpha$ it amounts to an inclusion of another minus sign
in the transformation of states under $z_3 \rightarrow -z_3$.
This has the effect that only twisted Ramond ground states that contribute to $H^{1,2}$ and
$H^{2,1}$ survive the projection. Thus we end up with a Hodge diamond mirror
to that of the orbifold without discrete torsion.
It was argued in \cite{VW} that these indeed constitute a mirror pair, where the mirror map
is T-duality.

Upon inclusion of a discrete torsion, the interesting supersymmetric cycles are the middle
dimensional ones. 
The construction of a boundary state is standard and we can follow the duality map.
There is, however, a delicate point.
The discrete torsion changes the projection operator, and for instance
in the $\alpha$ twisted sector it takes the form
\beq
P= \frac{1}{4}(1+\alpha-\beta-\gamma)
\comma
\eeq
which naively annihilates the twisted sector boundary state.
This is resolved by picking the correct ground state.
Consider the Ramond sector:
Related to $z_3$ we have the fermionic zero modes 
$\psi_{L,0}^3, \psi_{R,0}^3$
with the boundary condition
\beq
(\psi_{L,0}^3+i\eta \psi_{R,0}^{3*})|0\rangle = 0,~~~~~ 
(\psi_{L,0}^{3*}+i\eta \psi_{R,0}^3)|0\rangle = 0 
\comma
\eeq
with $\eta = \pm 1$. 

Of the possible Ramond ground states only 
$(i\eta \psi_{L,0}^3+\psi_{R,0}^3 + {\rm c.c})|0\rangle$ survives
the projection and should be picked.
This is to be contrasted with the case without discrete torsion where the 
correct twisted sector
Ramond ground state
is $(i\eta +\psi_{L,0}^3 \psi_{R,0}^3 + {\rm c.c})|0\rangle$.

Consider now the D-brane matrix  $R = {\sl diag}[\sigma_1,\sigma_1,\sigma_1]$.
A mirror symmetry transformation is defined by:
\beq
\pa z_i \rightarrow \pa z_i,~~~~~
\bar{\pa} z_i \rightarrow \bar{\pa} {\bar{z}_i}
\stop\eeq
Thus the matrix $T$ takes the form $T= {\sl diag} [\sigma_3,\sigma_3,\sigma_3]$ and 
does not commute with $R$. 
Since both $R$ and the mirror symmetry $T$ are equivariant with
respect
 to the $Z_2\times Z_2$ discrete
group, the same applies for the Calabi-Yau orbifold 
$(T^2)^3/(Z_2 \times Z_2)$, and we get the mixing
phenomena as we discussed before. 

In the orbifold models, we may consider gauge field strength
which belongs to the twisted sectors, namely localized on a particular
fixed point. In this case we should expect that the particular twisted
sector corresponding to this fixed point will be influenced. Thus, we are
led to consider different boundary conditions $R$ in (\ref{R}) for
the untwisted and twisted sectors. It would be interesting to further
explore this structure.

\section{Discussion}

We have shown that boundary states provide a framework at the SCFT level
to study configurations of D-branes, 
wrapped on supersymmetric cycles in Calabi-Yau spaces, with implications
on the structure of mirror symmetry between D-branes.
There are various directions for future research. 

Understanding the role of the non-chiral primary states in the 
boundary state is an important and challenging problem.
This is important since the non-chiral part carries information
on the moduli of supersymmetric cycles. 
One way to explore this issue is to use non-linear 
recursion relations for 
the boundary states which can be derived by moving vertex operators 
on the disc to its boundary and study the boundary states associated
with the disc splitting.

Boundary states can be used in order to explore
the moduli spaces of D-branes wrapped on supersymmetric cycles
in Calabi-Yau spaces. This will have various applications such as D-brane
states counting \cite{bsv2,vafa}, and
may provide us with means to probe the
structure of mirror symmetry as suggested in 
\cite{SYZ}. Moreover, we expect the boundary states to be also helpful
in exploring mirror symmetry in higher dimensions.
 
The relation that we found between open string worldsheet instanton
corrections and closed string instantons counting, points to a deep structure
between the corresponding moduli spaces which should be revealed.

It has been shown in \cite{DL} that supersymmetric gauge theories can be
realized via wrapping D-branes on supersymmetric cycles. The SCFT framework
that we presented is likely to be useful in exploring this direction.

\newpage
\section*{Acknowledgements}
We would like to thank K. ~Becker, T. ~Eguchi, D. ~Gepner, 
B. ~Greene, S.~Katz, M. ~Li, J. ~Maldacena, D. ~Morrison, 
R. ~Plesser, J. ~Polchinski, 
A. ~Schwimmer, A. ~Strominger and C. ~Vafa for useful discussions.
Y.O. would like to thank LBNL for hospitality during the final stages of
this work. This work was supported in part by the National Science
Foundation under grants PHS-9501018 and PHY-951497 and in part
by the Director, Office of Energy Research, Office of High Energy
and Nuclear Physics of the U.S. Department of Energy under Contract
DE-AC03-76SF00098. Y.O. is partially supported by the Israel Science Foundation through 
the Center for the Physics of Basic Interactions.
Z.Y. is supported by Graduate Research Fellowship
of the U.S. Department of Education. 

\newpage
\appendix{Boundary states for Gepner models}

A Gepner model \cite{Gepner} can be viewed as an orbifold construction
in which we project out states that do no satisfy the required conditions
and add twisted sectors to the Hilbert space.
This suggests that the way to construct the boundary state for a Gepner model
is to take the product of the boundary states for the minimal model parts
with the appropriate projection and addition of twisted sectors.

In the following we consider the simplest example: The $(k=1)^3$ Gepner model.
This corresponds to a sigma-model on $T^2$ with $Z_3$ symmetry. 
In this case, each minimal
model can be constructed by a free boson. Thus we have
$\phi_i$,~ $i=1,2,3$. Let us construct the boundary state for a D-brane 
wrapped on a supersymmetric 
1-cycle in $T^2$. Imposing the A-type boundary conditions implies 
\beq 
\phi_L^i = \phi_R^i + c_i
\comma
\label{cond}
\eeq
with constants $c_i$
\beq
c_i =  \frac{2\pi}{\sqrt{3}} n_i + (0 ~{\rm or}~ \frac{2\pi}{2
\sqrt{3}})
\comma \label{ci}
\eeq
where $n_i$ are integers and the choice of $0$ or 
 $ \frac{2\pi}{2 \sqrt{3}}$ corresponds to the sign of 
the Ramond-Ramond charge (i.e. BPS or anti-BPS). 
For each choice of $c_i$, the boundary state is uniquely
constructed by the standard oscillator procedure.

It is instructive to interpret this from the
sigma-model viewpoint. 
The sigma-model for $T^2$ consists
of complex free boson $X$ and a complex free fermion $\psi$ which are related
to $\phi_i$ by 
\beqar
\psi &=& {\rm exp} \left[ \frac{i}{\sqrt{3}}(\phi_1+\phi_2+\phi_3)
\right]\comma
\nonumber\\
\pa X  &=& {\rm exp} \left[ \frac{i}{\sqrt{3}} (-2\phi_1
+\phi_2+\phi_3)\right]~ +~
({\rm permutations~ in}~1,2,3)
\stop
\eeqar
The boundary conditions  (\ref{cond}),(\ref{ci}) correspond in the sigma model
to
\beqar
\psi_L &=& \pm e^{ \frac{2\pi i}{3} (n_1+n_2+n_3) }
\psi_R
\comma
\nonumber\\
\pa X &=& e^{\frac{2\pi i}{3} (n_1+n_2+n_3)} \bar{\pa} X
\stop
\eeqar

The case $n_1+n_2+n_3 = 0~{\rm mod}~ 3$ corresponds to the Neumann 
boundary condition on the $\{X= {\rm real}\}$ cycle of $T^2$, while
$ n_1+n_2+n_3 = 1$ or $2$ ${\rm mod}~3$ correspond 
to Neumann boundary conditions on the $Z_3$ related 1-cycles.
We see that the different choices
of boundary conditions for the Gepner model correspond
to the different choices of supersymmetric 1-cycles.
We expect that such relations between the algebraic 
and the geometric structures should exist in general.

The boundary state takes the form
$|B\rangle = |B\rangle_X |B\rangle_{\psi}$ where
\beqar
|B\rangle_X &=& {\rm exp} \left[-e^{\frac{2\pi i}{3} (n_1+n_2+n_3)}
(\sum_{n=1}^{\infty}\frac{1}{n}\alpha_{L,-n}
\alpha_{R,-n} + {\rm c.c} )\right]|0\rangle,
\nonumber\\
|B\rangle_{\psi} &=& {\rm exp}\left[\pm i e^{\frac{2\pi i}{3} (n_1+n_2+n_3)}
(\sum_{n}\psi_{L, -n}
\psi_{R, -n} + {\rm c.c}) \right]|0\rangle
\stop
\label{bt2}
\eeqar
Note that from the chiral primary states only 
the $(c,c)$ ring 
$\{1,~  
\psi_L\psi_R\}$
and its complex conjugate $(a,a)$ ring contribute to the boundary state
as expected.


\newpage
 
\begin{thebibliography}{99}

\small
\parskip=0pt plus 2pt
\bibitem{POL} J.~Polchinski,
``Dirichlet-Branes and Ramond-Ramond Charges,'' hep-th 9510017,
\prl 75,95,4724.
\bibitem{bsv2} M.~Bershadsky, V.~Sadov and C.~Vafa, ``D-Branes and 
Topological Field Theories,''
hep-th 9511222, \np463,96,420.
\bibitem{boundary}  J.~Polchinski and Y.~Cai,
``Consistency of Open Superstring
Theories,'' \np296,88,91;
C.~G.~Callan, C.~Lovelace, C.~R.~Nappi and 
S.~A.~Yost, ``Loop Corrections to Superstring Equations of Motion,''
\np308,88,221.
\bibitem{al} P.~S.~Aspinwall and C.~A.~L\"utken, ``Quantum Algebraic
Geometry of Superstring Compactifications,'' \np355,91,482.
\bibitem{AM} P.~S.~ Aspinwall and D.~R.~Morrison, 
``U-Duality and Integral Structures,'' hep-th 9505025, \pl355,95,141.
\bibitem{BBS} K.~Becker, M.~Becker and A.~Strominger, ``Fivebranes,
Membranes and Nonperturbative String Theory,''
hep-th 9507158, \np456,95,130.
\bibitem{BV} N.~Berkovits and C.~Vafa, ``$N=4$ Topological Strings,''
hep-th 9407190, \np433,95,123.
\bibitem{Joyce} D.~Joyce, ``Compact 8-Manifolds with Holonomy
$Spin(7)$,'' \invm123,96,507. 
\bibitem{OOZ} H.~Ooguri, Y.~Oz and Z.~Yin, work in progress.
\bibitem{Ishibashi} N.~Ishibashi, ``The Boundary and Crosscap States in 
Conformal Field Theories,'' 
\mpl4,89,251.
\bibitem{Strominger} A.~Strominger, ``Special Geometry,''
\cmp133,90,163.
\bibitem{Candelas} P.~Candelas and X.~C.~de la Ossa,
``Moduli Space of Calabi-Yau 
Manifolds,'' \np355,91,455.
\bibitem{CV} S.~Cecotti and C.~Vafa, ``Topological Antitopological
Fusion,'' \np367,91,359.
\bibitem{BCOV} M.~Bershadsky, S.~Cecotti, H.~Ooguri and C.~Vafa,
``Kodaira-Spencer Theory of Gravity and Exact Results for
Quantum String Amplitudes,'' hep-th 9309140, \cmp165,94,311.
\bibitem{GMP} B.~R.~Greene, D.~R.~Morrison and M.~R.~Plesser,
``Mirror Manifolds in Higher Dimension,''
hep-th 9402119, \cmp173,95,559.
\bibitem{GP} B.~R.~Greene and M.~R.~Plesser,
``Duality in Calabi-Yau Moduli Space,'' \np338,90,15.
\bibitem{Morrison} D.~R.~Morrison, ``Where is the Large Radius Limit?,''
 hep-th 9311049, Proceedings of Strings 93. 
\bibitem{cogp} P.~Candelas, X.~C.~De\thinspace La\thinspace Ossa,
P.~S.~Green and P.~Parkes, ``A Pair of Calabi-Yau Manifolds as an Exactly Solubule
Superconformal Theory,'' \np359,91,21.
\bibitem{morrison3} D.~Morrison, ``Mirror Symmetry and Rational Curves
on Quintic Threefolds: A Guide for Mathematicians,'' \jams6,93,223.
\bibitem{Schmid} W.~Schmid, ``Variation of Hodge Structure:
The Singularities of the Period Mapping,'' \invm22,73,211. 
\bibitem{Morrison1} D.~R.~Morrison, ``Making Enumerative Predictions by
Means of Mirror Symmetry,'' alg-geom 9504013, Essays on Mirror Manifolds II.
\bibitem{SYZ} A.~Strominger, S.-T.~Yau and E.~Zaslow,
``Mirror Symmetry is T-Duality," hep-th 9606040.
\bibitem{Morrison2} D.~R.~Morrison,
``Mirror Symmetry and the Type II String,'' hep-th 9512016.
\bibitem{Witten} E.~Witten, ``Bound States of Strings and p-Branes,"
hep-th 9510135, \np460,96,335.  
\bibitem{Douglas} M.~R.~Douglas,
``Branes within Branes,'' hep-th 9512077.
\bibitem{VW} C.~Vafa and E.~Witten, ``On Orbifolds with Discrete Torsion,"
hep-th 9409188, \jmphy15,95,189.
\bibitem{vafa} C.~Vafa, ``Instantons on D-branes,''
hep-th 9512078, \np463,96,435.
\bibitem{DL} M.~R.~Douglas and M.~Li,
``D-Brane Realization of $N=2$ Super Yang-Mills Theory in Four
Dimensions,'' hep-th 9604041. 
\bibitem{Gepner} D.~Gepner, 
``Exactly Solvable String Compactifications on Manifolds of $SU(N)$
Holonomy,'' \pl199,87,380. 
    





\end{thebibliography}
\end{document}